\documentclass[runningheads]{llncs}
\usepackage[T1]{fontenc}
\usepackage{graphicx}
\usepackage{amsmath}
\usepackage{float} 
\usepackage{placeins} 
\usepackage{subcaption} 
\usepackage{microtype}
\usepackage{hyperref}
\usepackage{color}

\begin{document}
\title{Unstable Prompts, Unreliable Segmentations: \newline A Challenge for Longitudinal Lesion Analysis}
\titlerunning{Unstable Prompts, Unreliable Segmentations}
\author{Niels Rocholl\inst{1} \and
Ewoud Smit\inst{1} \and
Mathias Prokop\inst{1} \and
Alessa Hering\inst{1}}
\authorrunning{N.M.M. Rocholl et al.}
\institute{Department of Medical Imaging, Radboudumc, Nijmegen, The Netherlands \\
\email{niels.rocholl@radboudumc.nl}}
\maketitle              %
\begin{abstract}
Longitudinal lesion analysis is crucial for oncological care, yet automated tools often struggle with temporal consistency. While universal lesion segmentation models have advanced, they are typically designed for single time points. This paper investigates the performance of the ULS23 segmentation model in a longitudinal context. Using a public clinical dataset of baseline and follow-up CT scans, we evaluated the model's ability to segment and track lesions over time. We identified two critical, interconnected failure modes: a sharp degradation in segmentation quality in follow-up cases due to inter-scan registration errors, and a subsequent breakdown of the lesion correspondence process. To systematically probe this vulnerability, we conducted a controlled experiment where we artificially displaced the input volume relative to the true lesion center. Our results demonstrate that the model's performance is highly dependent on its assumption of a centered lesion; segmentation accuracy collapses when the lesion is sufficiently displaced. Our results reveal a fundamental limitation of applying single-timepoint models to longitudinal data. We conclude that robust oncological tracking requires a paradigm shift away from cascading single-purpose tools towards integrated, end-to-end models that are inherently designed for temporal analysis.

\keywords{Longitudinal Analysis \and Lesion Segmentation \and Computed Tomography \and Deep Learning \and Registration Error.}
\end{abstract}
\section{Introduction}
\label{sec:intro}

Longitudinal imaging is essential for monitoring disease progression and treatment response in oncology. With the number of CT examinations steadily increasing and the global cancer burden projected to rise by 47\% by 2040 \cite{GLOBOCAN_Sung_2021}, radiologists face growing pressure to manage high imaging workloads effectively. Cancer patients often undergo multiple CT scans during treatment and follow-up, and response evaluation guidelines such as RECIST 1.1 \cite{RECIST_eisenhauer_2009} are frequently applied to measure lesion changes across time. However, lesion measurement and response assessment are still predominantly manual, requiring significant reading time and subject to inter-observer variability.
Automatic lesion segmentation models offer the potential to reduce this burden by enabling consistent and efficient lesion tracking. Recent studies \cite{ImprovingAssessmentLesions_hering_2024,AssistedManualInterpretation_jacobs_2021,RandomizedMultireaderEvaluation_lu_2021} have demonstrated that automation of follow-up evaluation can reduce both reading times and disagreement among readers.

While the Universal Lesion Segmentation (ULS \cite{ULS23_grauw_2024}) dataset enabled the development of general-purpose lesion segmentation models across a wide range of lesion types, it remains limited to single time points and lacks temporal context. To address this, a new publicly available longitudinal CT dataset \cite{kustner_LongitudinalCT_2025} was recently introduced, comprising baseline and follow-up scans from melanoma patients undergoing systemic therapy. The dataset includes manual annotations of all visible metastases per patient, offering a valuable benchmark for evaluating lesion-tracking algorithms in longitudinal imaging.

This paper presents an in-depth evaluation of the baseline model from the ULS23 challenge on longitudinal CT scans with annotated metastases \cite{ULS23_grauw_2024}. We identify two recurrent and clinically relevant failure modes: (1) spatial drift in lesion localization across timepoints caused by inter-scan registration errors, and (2) the misidentification or complete loss of lesions in follow-up scans as a result. We formalize these issues, quantify their occurrence, and illustrate their impact on lesion tracking. Our findings offer practical insights that can guide the development of more robust and temporally consistent lesion segmentation models.

\section{Materials and methods}
\label{sec:methods}
To systematically evaluate the ULS23 baseline model in a longitudinal context, we designed a two-part experimental framework. The first experiment assesses the model's overall performance on the publicly available longitudinal CT dataset \cite{kustner_LongitudinalCT_2025}, focusing on both baseline and follow-up scans. The second, more targeted experiment, investigates the influence of registration errors of varying magnitudes on the segmentation output.

\subsection{ULS23 segmentation model}
\label{ssec:uls23_model}
The ULS23 model \cite{ULS23_grauw_2024} is a semi-supervised 3D universal lesion segmentation system, developed within the nnU-Net v1 framework \cite{nnUNet_isensee_2021}. It employs a 3D Residual Encoder UNet architecture, which was extended from the standard nnU-Net configuration with additional encoder and decoder layers (for a total of seven) and an increased feature channel capacity (upper bound raised from 320 to 384). 

A defining characteristic of ULS23 is that it disables the default resampling and input patching strategies typically used in nnU-Net, in order to preserve spatial priors in the input data. Consequently, ULS23 operates on fixed-size Volumes of Interest (VOIs) of $256 \times 256 \times 128$ voxels and, assumes that a single target lesion is located at the center of each VOI. These VOIs are typically generated by cropping around suspected lesion locations.

The training of ULS23 involved a two-stage, semi-supervised process. The model was initially pre-trained using 2D pseudo-masks generated from partially annotated datasets and subsequently fine-tuned on a large, aggregated collection of 3D ground-truth masks. In the second stage, this model was used to generate improved, quality-filtered 3D pseudo-labels. A final fine-tuning step was then performed exclusively on the high-quality 3D ground-truth data, which was derived from VOIs centered on the target lesions. This training strategy, particularly its reliance on accurately centered lesions in VOIs, is a key aspect considered in our analysis of its application to longitudinal data.

\subsection{Longitudinal CT dataset}
\label{ssec:longitudinal_dataset}
All 300 scans were acquired at the University Hospital Tübingen (UKT) as part of routine treatment assessment in patients with metastatic melanoma. Each case includes manual 3D segmentations of all malignant lesions, both primary tumors and metastases, at both time points. For each lesion identified on the baseline scan, a coordinate is provided at the center of the lesion (centroid, i.e., the geometric center point). In the corresponding follow-up scan, both the true lesion centroid and a centroid propagated from baseline via conventional image registration are provided (note that propagated centroids are not available for all lesions in the dataset). All images were acquired on CT systems (Siemens Healthineers, Erlangen, Germany)  using a standardized whole-body staging protocol during the portal-venous contrast phase, with a reconstructed slice thickness of 3 mm. Crucially, this dataset is publicly available, ensuring our findings are fully reproducible.

\begin{figure}[htb!]
    \centering
    \includegraphics[width=0.9\linewidth]{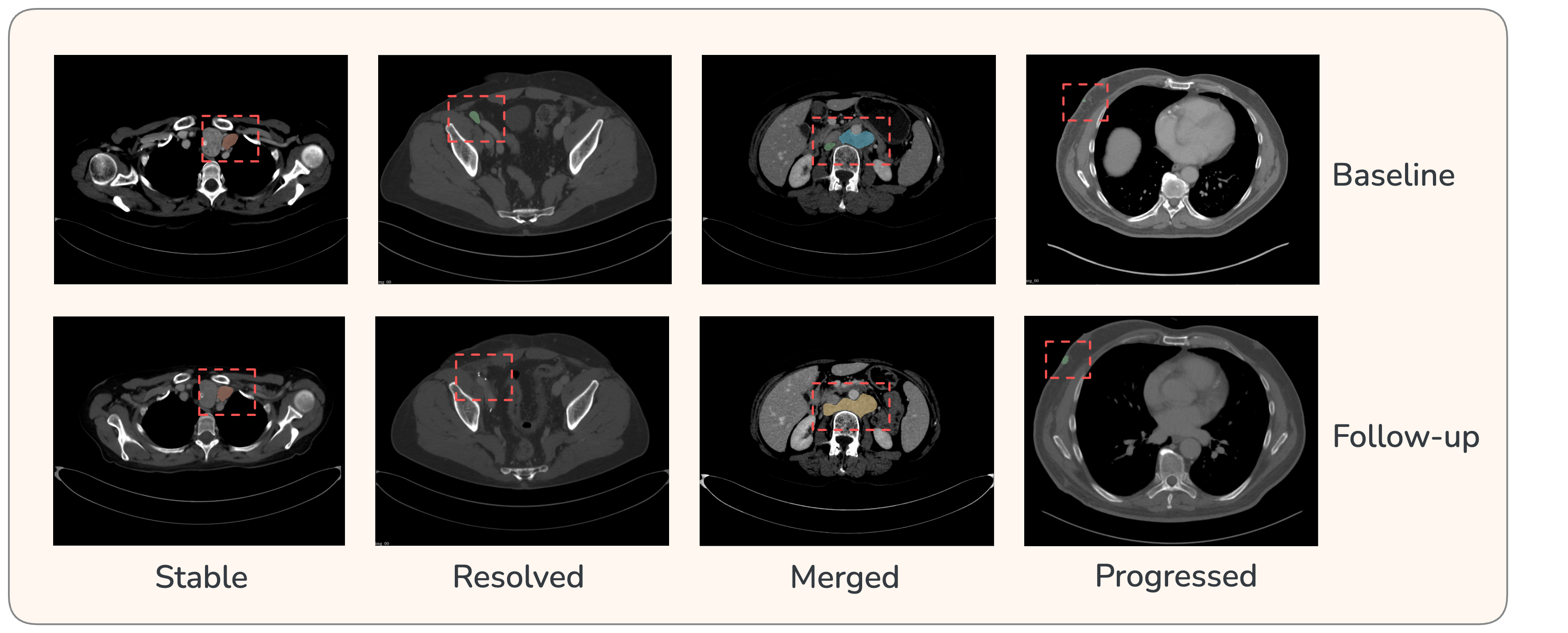}
    \caption{Four examples from the public Longitudinal CT dataset, showing stable, resolved, merged, and progressed lesions.}
    \label{fig:longitudinal_ct_examples}
\end{figure}

The dataset captures a broad range of longitudinal lesion changes, including growth, shrinkage, splitting, merging, complete disappearance, and the appearance of new lesions. Four representative examples from the dataset are shown in Figure \ref{fig:longitudinal_ct_examples}, illustrating different lesion progression patterns such as stable, resolved, merged, and progressed lesions.

\subsection{Experimental setup and evaluation}
\label{ssec:experimental_setup}
We designed two experimental protocols to evaluate the ULS23 model in a longitudinal setting. The first assesses its performance on both baseline and follow-up scans. The second experiment quantifies the model's dependence on its centered-lesion assumption by systematically introducing controlled misalignments and measuring the impact on segmentation performance. The overall workflow of both experiments is visualized in Figure \ref{fig:experiment_workflows}.

\subsubsection{Longitudinal performance evaluation} %
\label{sssec:exp_longitudinal_eval}

\begin{figure}[htb!]
    \centering
    \begin{subfigure}{\linewidth}
        \centering
        \includegraphics[width=\linewidth]{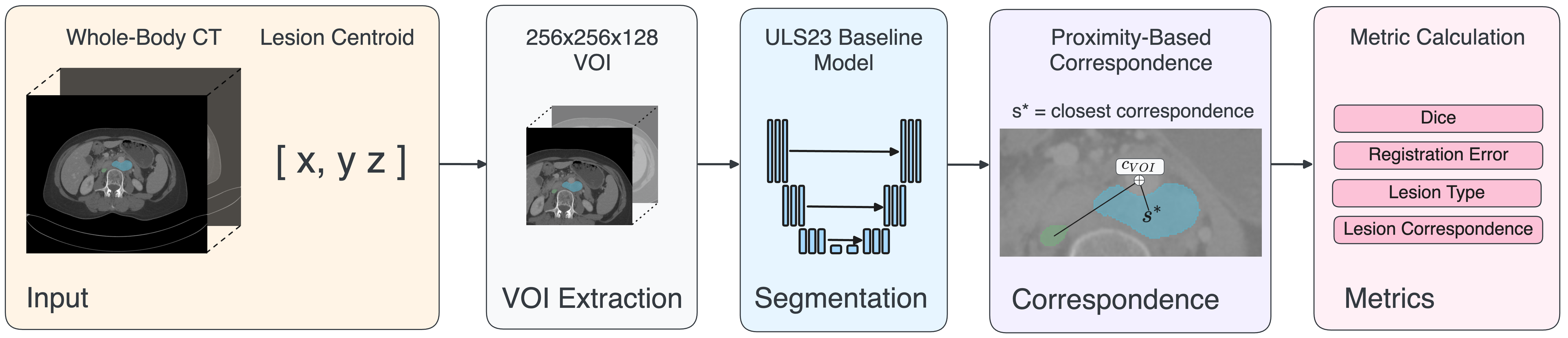}
        \caption{Workflow for the Longitudinal performance evaluation experiment. VOIs are extracted around lesion centroids (propagated for follow-up), segmented by ULS23, correspondence is determined, and metrics are logged.}
        \label{fig:workflow_general}
    \end{subfigure}
    
    \hfill
    
    \begin{subfigure}{\linewidth}
        \centering
        \includegraphics[width=\linewidth]{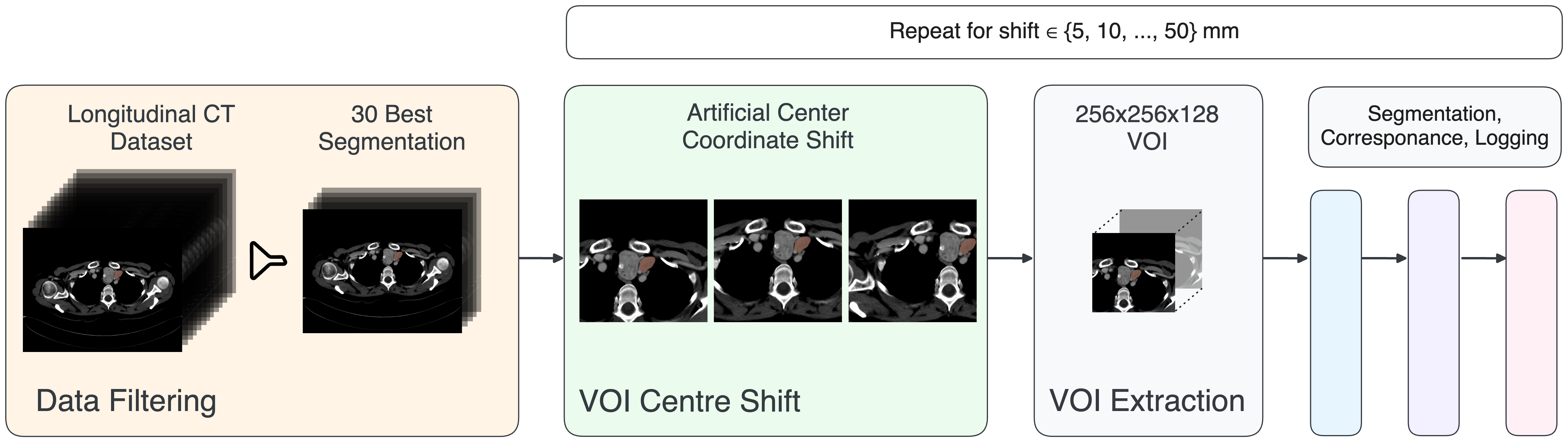}
        \caption{Workflow for the Controlled VOI Displacement Analysis. VOIs from the 30 best-segmented lesions are systematically shifted before segmentation to quantify the impact.}
        \label{fig:workflow_shift}
    \end{subfigure}

    \caption{Schematic overview of the two experimental setups: (a) evaluating ULS23 performance on the full longitudinal dataset, and (b) analyzing ULS23 sensitivity to controlled VOI center displacements.}
    \label{fig:experiment_workflows}
\end{figure}

The first experiment follows a relatively standard four-step segmentation pipeline: VOI extraction, segmentation, lesion correspondence, and metric calculation. This process is schematically outlined in Figure \ref{fig:experiment_workflows}. For a single CT scan, the process unfolds as follows: we iterate through the provided lesion coordinates of the CT volume, each associated with a unique lesion ID. A $256 \times 256 \times 128$ voxel VOI was extracted for each coordinate. For baseline scans, VOIs were centered on the given lesion centroids. For follow-up scans, VOIs were centered on the propagated baseline centroids when available, otherwise, the true follow-up lesion centroid was used as a best-case scenario reference. While centroids for the baseline scans can be assumed to be at the true center of gravity of each lesion, follow-up centroids propagated through registration can be subject to registration error. Inference was then performed on each extracted VOI, yielding a 3D segmentation mask. To establish lesion correspondence, we first identified all distinct connected components within the raw segmentation output. The lesion ID of the current VOI was then assigned to the single connected component whose centroid was closest to the VOI's geometric center. Formally we say that if $S = \{s_1, s_2, \ldots, s_n\}$ is the set of $n$ segmented connected components, $c(s_i)$ is the centroid of component $s_i$, and $c_{\text{VOI}}$ are the coordinates of the VOI center, the assigned component $s^*$ is determined by:
$$s^* = \arg\min_{s_i \in S} \|c(s_i) - c_{\text{VOI}}\|_2$$
As each VOI is processed with the expectation of segmenting a single central lesion, any other segmented components within that VOI were not considered for the metrics associated with that specific lesion.

We defined the outcome of each VOI-level segmentation based on whether the selected component $s^*$ correctly corresponded to the same lesion instance as identified in the baseline scan. This categorization accounts for both successful tracking and the possibility of lesion resolution:
\begin{itemize}
    \item \textbf{Correct Assignment:} The selected component $s^*$ corresponds to the same lesion as identified in the baseline scan. This includes both anatomically and temporally correct matches.
    \item \textbf{True Negative (TN):} No segmentation was produced, and the baseline lesion had resolved, meaning no follow-up lesion was present.
    \item \textbf{Incorrect Assignment:} The component $s^*$ does not correspond to the same lesion as identified in the baseline scan. This includes two common cases:
A different lesion was segmented (i.e., wrong match within the VOI), or a segmentation was produced despite the baseline lesion having resolved (i.e., the lesion was no longer present but the model hallucinated one).
    \item \textbf{False Negative (FN):} No component was segmented as $s^*$, despite the presence of a ground truth target lesion.
\end{itemize}

For evaluation, we calculated the Dice similarity coefficient for each segmented lesion against its corresponding ground truth. To assess the impact of temporal changes, we compared the distribution of Dice scores between baseline and follow-up scans using the Wilcoxon signed-rank test. 

\subsubsection{Controlled VOI displacement analysis}
\label{sssec:exp_displacement_analysis}
The second experiment was designed to systematically evaluate the ULS23 model's sensitivity to one of its core assumptions: that lesions are located at the center of the input VOI. The experimental workflow is illustrated in Figure \ref{fig:workflow_shift}. Initial analyses of the longitudinal data (Section \ref{sssec:exp_longitudinal_eval}) indicated that while registration errors often led to off-center lesions, the naturally occurring instances were insufficient for a statistically robust analysis of this specific effect across a controlled range of misalignments. Furthermore, the inherent variability in ULS23's baseline performance on diverse lesions could confound the impact of VOI misalignments.

To create a reliable reference set for the displacement study, we selected the 30 lesions that achieved the highest Dice in the longitudinal-performance experiment. These well-segmented cases provide a consistent, high-quality baseline against which the impact of controlled VOI shifts can be measured more precisely.

For each of these 30 selected lesions, the VOI's designated center coordinate was then artificially shifted from the true lesion centroid. To ensure shifts do not move into zero padded space, they were applied along a direction unique to each lesion, defined by the vector from its true centroid to its CT volume center. These perturbations were applied using magnitudes $\epsilon \in \{0, 5, 10, 15, 20, 25, 30, 35, 40, 45, 50\}$ mm. The $\epsilon=0$ mm case represents the original, accurately centered VOI for these top-performing lesions. For each resulting shifted coordinate ($c_{\text{shifted}}$), a $256 \times 256 \times 128$ voxel VOI was extracted, centered at $c_{\text{shifted}}$. Subsequently, the ULS23 model performed segmentation, followed by the same lesion correspondence and metric calculation procedures detailed in Section \ref{sssec:exp_longitudinal_eval}. This allowed for a direct comparison of segmentation quality and correspondence accuracy as a function of controlled VOI displacement from the true lesion center.

\section{Results}
\label{sec:results}

\subsection{Longitudinal Performance Evaluation}
\label{ssec:longitudinal_performance}

\begin{figure}[htb!]
    \centering
    \begin{subfigure}[b]{0.48\linewidth}
        \centering
        \includegraphics[width=\linewidth]{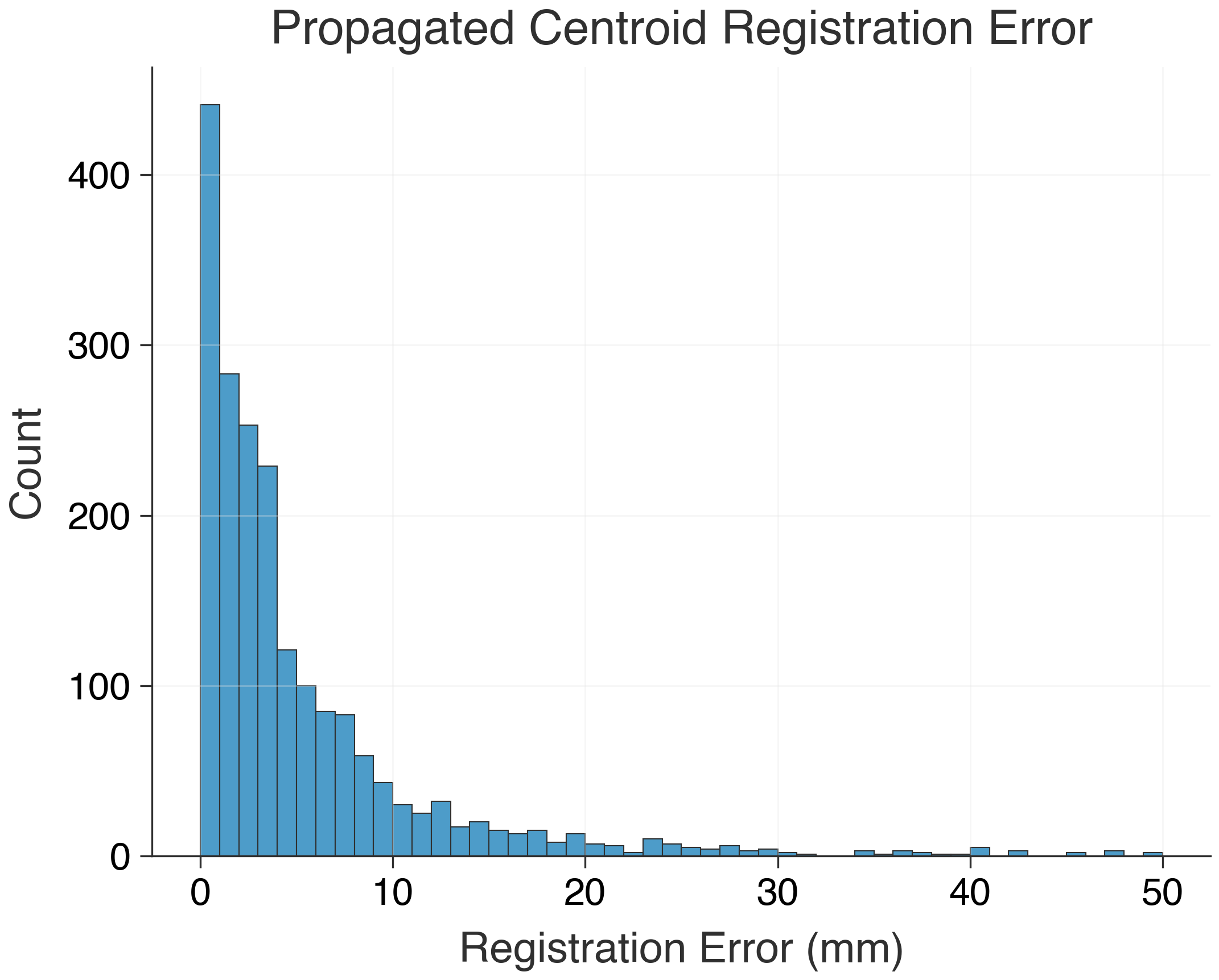}
        \caption{}
        \label{fig:reg_error_dist}
    \end{subfigure}
    \hfill
    \begin{subfigure}[b]{0.48\linewidth}
        \centering
        \includegraphics[width=\linewidth]{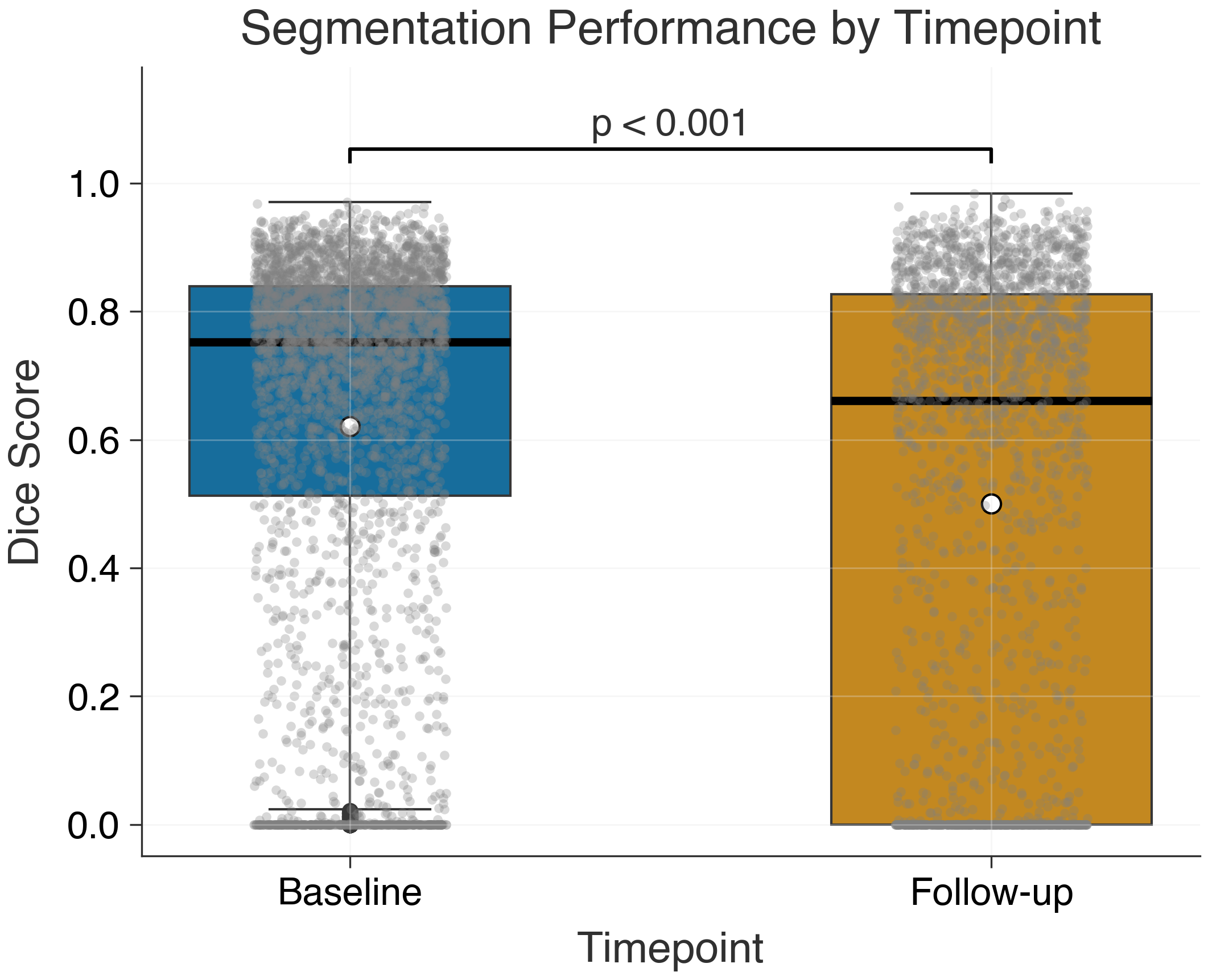}
        \caption{}
        \label{fig:bl_vs_fu_dice}
    \end{subfigure}
    
    \caption{The impact of registration error on longitudinal segmentation performance. (a) The distribution of registration errors for propagated centroids in follow-up scans reveals a long tail of significant misalignments, with a notable number of lesions displaced by over 10 mm. (b) These errors lead to a substantial drop in Dice score at follow-up compared to baseline (p < 0.001), indicating frequent segmentation failures.}
    \label{fig:longitudinal_performance_plots}
\end{figure}

Figure \ref{fig:reg_error_dist} highlights the distribution of registration errors for all propagated lesion coordinates. These are calculated by taking the Euclidean distance from the propagated centroid to the centroid of the ground truth mask. Each bar comprises a 1 mm range bin. The histogram shows us that while a substantial portion of propagated centroids has a low registration error, the distribution has a long tail, with a non-trivial number of lesions exhibiting errors greater than 10 mm.

This registration error directly translates to poorer segmentation performance at the follow-up time point. As shown in Figure \ref{fig:bl_vs_fu_dice}, the Dice is significantly lower for the follow-up scans compared to the baseline scans (p < 0.001). The distribution of Dice scores for the follow-up scans is also more dispersed, with a larger proportion of very low scores, indicating frequent and severe segmentation failures.

\begin{figure}[htb!]
    \centering
    \begin{subfigure}[b]{0.48\linewidth}
        \centering
        \includegraphics[width=\linewidth]{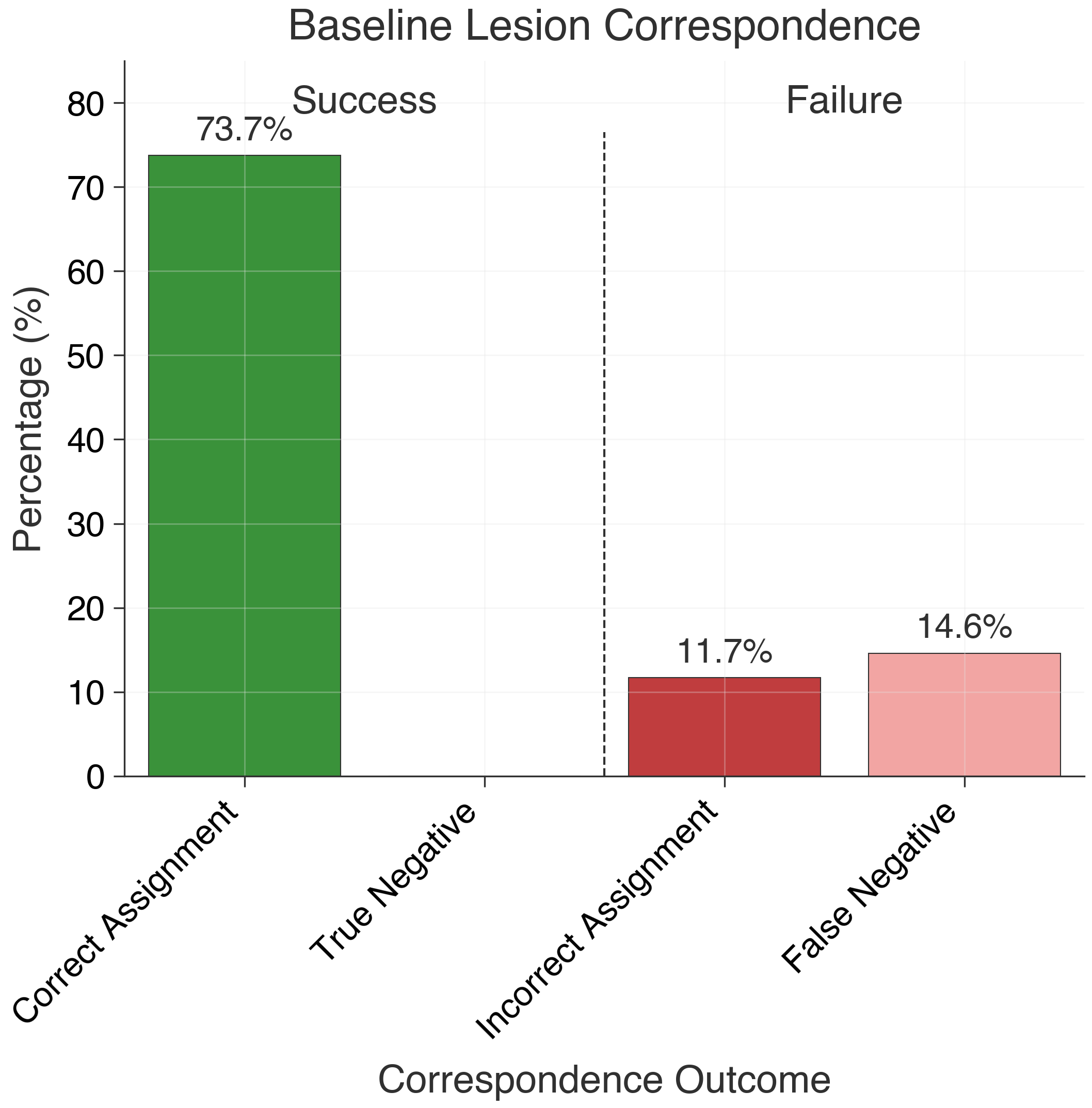}
        \caption{}
        \label{fig:lesion_assignment_a}
    \end{subfigure}
    \hfill
    \begin{subfigure}[b]{0.48\linewidth}
        \centering
        \includegraphics[width=\linewidth]{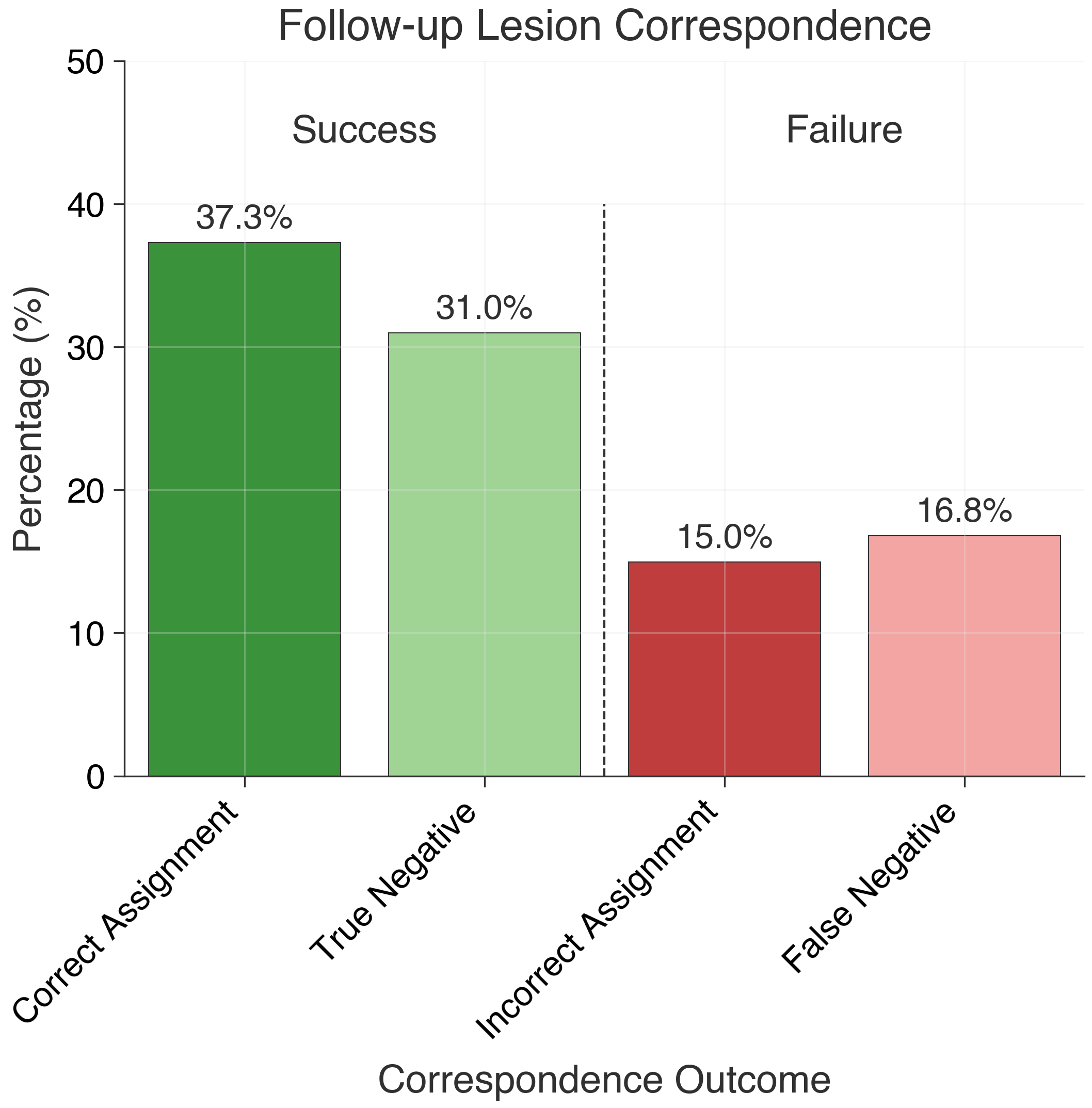}
        \caption{}
        \label{fig:lesion_assignment_b}
    \end{subfigure}
    
    \caption{Lesion correspondence outcomes for (a) baseline and (b) follow-up scans. The rate of clinically critical tracking errors is high at both time points: the combined rate of incorrect assignments and false negatives is 26.3\% at baseline and worsens to 31.8\% at follow-up. This highlights the fragility of the correspondence method when segmentation is unreliable, a direct consequence of the issues shown in Figure \ref{fig:longitudinal_performance_plots}.}

    \label{fig:lesion_assignment_plots}
\end{figure}

Figure \ref{fig:lesion_assignment_a} and \ref{fig:lesion_assignment_b} illustrate the results of our lesion correspondence method. For the baseline cases, we can see that in the majority of cases, the correct lesion is assigned, with some incorrect assignments, as well as some false negatives. 
In contrast, the follow-up results show a substantial decrease in correct assignments and a notable increase in incorrect assignments, false negatives, and, most notably, true negatives. The number of true negatives (where the model correctly segments nothing) increases from zero at baseline to 31\% at follow-up. 

\subsection{Controlled VOI Displacement Analysis}

\begin{figure}[htb!]
    \centering
    \begin{subfigure}[b]{0.48\linewidth}
        \centering
        \includegraphics[width=\linewidth]{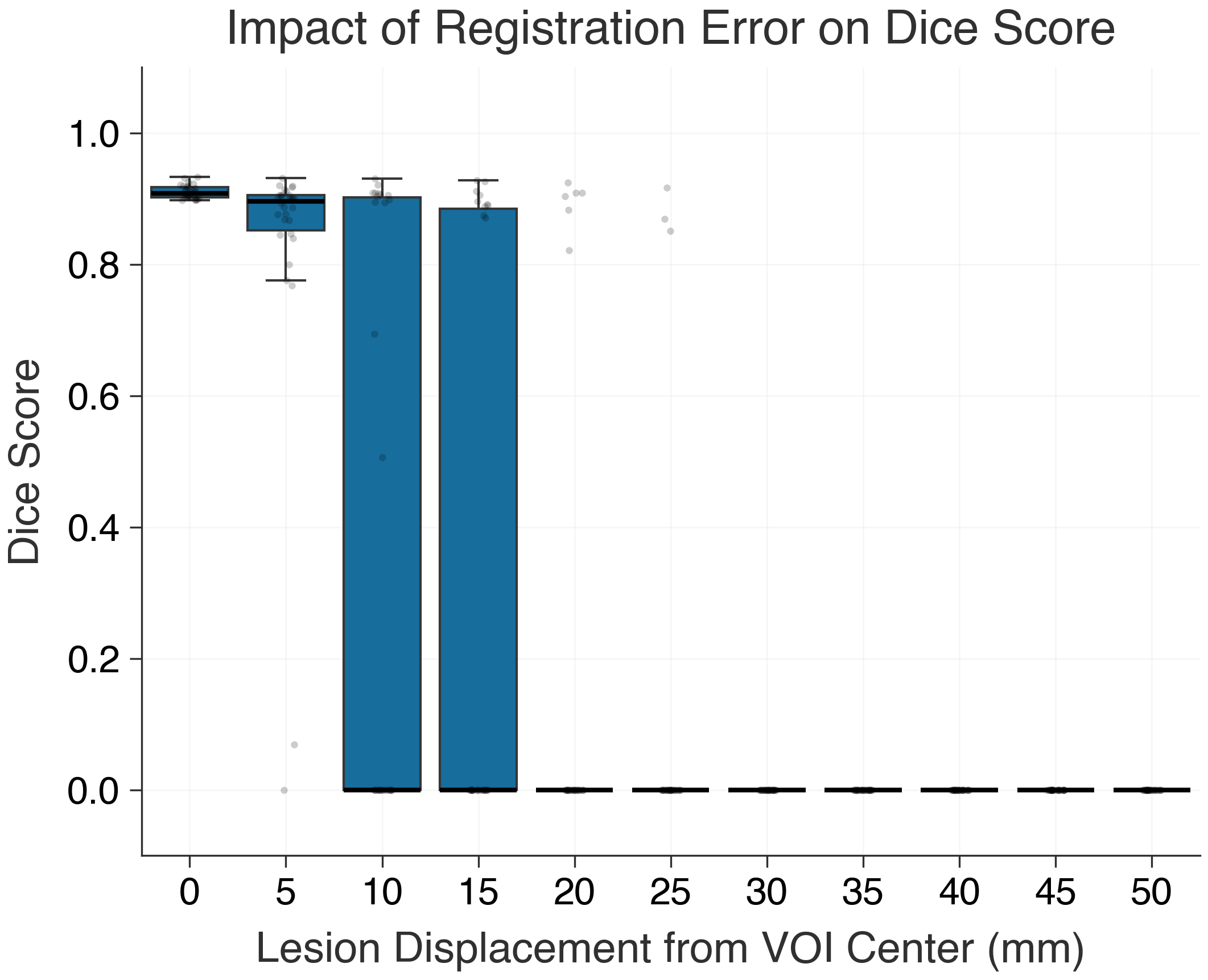}
        \caption{}
        \label{fig:dice_vs_distance}
    \end{subfigure}
    \hfill
    \begin{subfigure}[b]{0.48\linewidth}
        \centering
        \includegraphics[width=\linewidth]{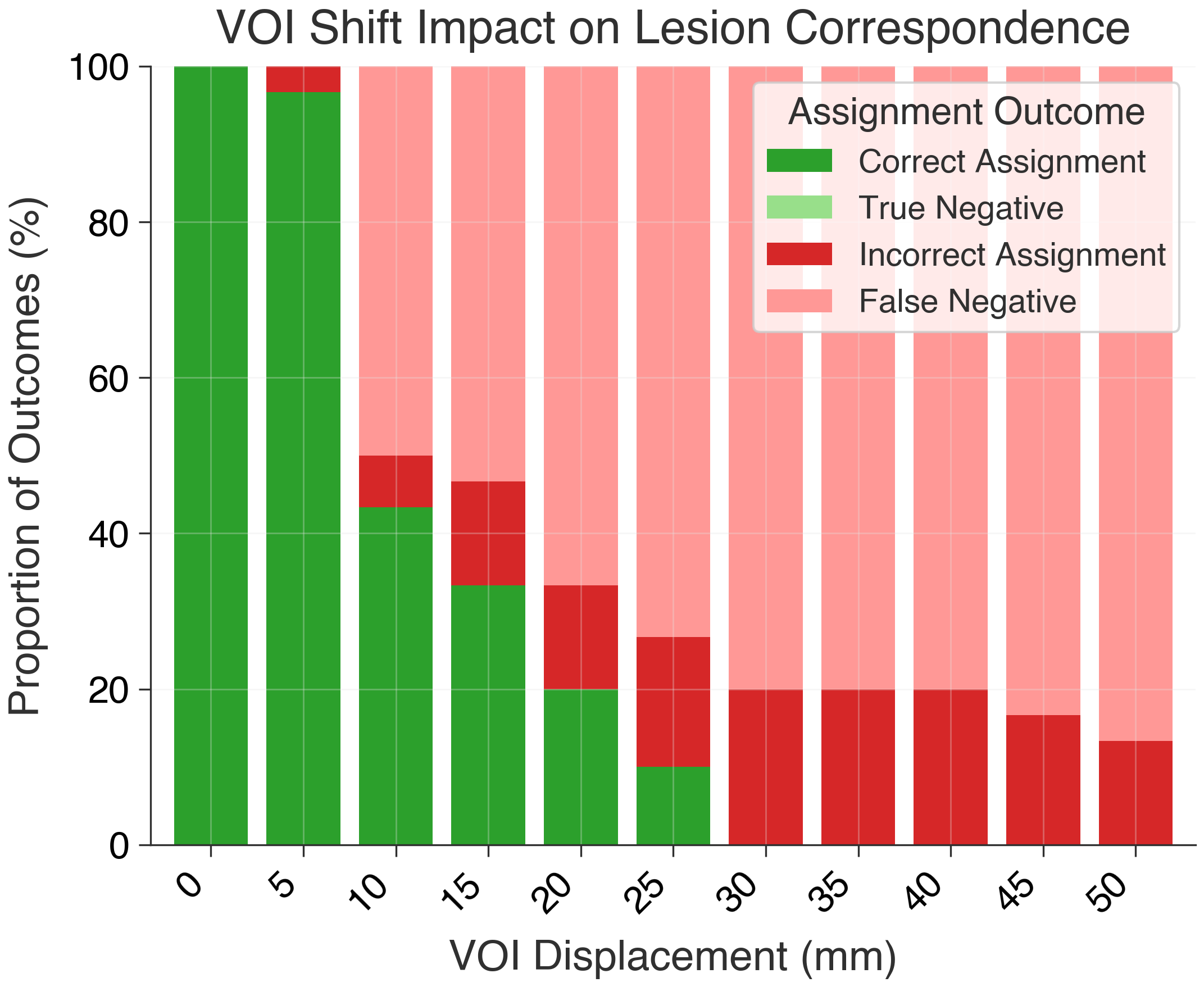}
        \caption{}
        \label{fig:correspondence_impact_shift}
    \end{subfigure}
    
    \caption{Controlled experiment showing ULS23's sensitivity to lesion displacement. (a) The Dice score shows a sharp drop when the lesion is displaced more than ~20mm from the VOI center, indicating complete segmentation failure. (b) Consequently, the rate of correct lesion correspondence falls drastically, with failures dominating even at moderate (10-15 mm) displacements. This controlled experiment confirms that the performance drop seen in real-world follow-up data (Figure \ref{fig:longitudinal_performance_plots}) is caused by VOI misalignment.}
    \label{fig:displacement_analysis_plots}
\end{figure}

Figure \ref{fig:displacement_analysis_plots} shows the results of the controlled VOI displacement experiment. Here the true centroid coordinate of the 30 top segmented lesions was systematically shifted, to simulate registration error. 

Figure \ref{fig:dice_vs_distance} shows how segmentation performance degrades with increasing (synthetic) registration errors. Instead of a gradual decline, the relationship between the Dice score and the lesion-to-VOI-center distance is defined by a sharp threshold. The model's ability to accurately segment the lesion is high for displacements below approximately 20 mm but deteriorates rapidly beyond that point. For these larger displacements, the Dice score is almost universally near zero, indicating a complete segmentation failure in which the predicted mask shares virtually no mutual information with the ground truth. 

Figure \ref{fig:correspondence_impact_shift} illustrates the proportion of lesion correspondence outcomes as a function of artificially introduced displacement. At 0 mm displacement, 100\% of assignments are correct. However, the proportion of correct assignments rapidly declines with increasing displacement, falling below 50\% by 10 mm. Correspondingly, incorrect assignments and false negatives become more frequent. Beyond a 25 mm shift, there are virtually no correct assignments, with false negatives (i.e., complete failure to segment anything) becoming the dominant outcome. 

\section{Discussion}
\label{sec:discussion}
The experimental results reveal two fundamental challenges when applying the ULS23 baseline model in a longitudinal context. The first is the model's sensitivity to VOI misalignment, stemming from the violation of its assumption that lesions are always centered in the VOI. The ULS23 model operates under a strong assumption inherited from its training: the lesion is always located at the center of the VOI. Additionally, this bias was reinforced by its training on non-exhaustively annotated data, teaching the model to disregard peripheral lesions. 

In a longitudinal setting, this core assumption breaks down. Propagated centroid can suffer from registration errors that frequently displace the target lesion from the VOI's center. This misalignment can result in a significant degradation of segmentation quality, evidenced by the sharp drop in Dice scores from the baseline to the follow-up scans (Figure \ref{fig:bl_vs_fu_dice}). It is important, however, to consider confounding factors beyond geometric misalignment. For instance, changes in lesion morphology and appearance between scans, resulting from treatment or disease progression, are confounding factors that could also contribute to the observed performance degradation. This, combined with the appearance of new and disappearance of old lesions, complicates a direct one-to-one comparison between baseline and follow-up performance.

The controlled experiment detailed in Section \ref{sssec:exp_displacement_analysis} provided a systematic analysis of how VOI-to-lesion displacement impacts segmentation performance. The results, illustrated in Figure \ref{fig:dice_vs_distance}, demonstrate a clear and critical relationship: while segmentation quality (Dice score) is robust to minor shifts, it degrades sharply as the lesion becomes more off-center, often failing completely once a distance threshold is exceeded. This behavior can be naturally explained by the ULS23 model's training assumption to segment the central object, disregarding peripheral structures as background. It is worth noting that some of the larger, artificial displacements may not be realistic for a typical registration algorithm (e.g., shifting a coordinate from one organ to another). However, the results still imply a fundamental principle: the model's success is highly contingent on the accuracy of the guiding coordinate. If this centroid, whether sourced from an automated registration or a manual annotation such as a radiologist's click, is sufficiently far from the actual lesion, the model is designed to ignore it.

The second challenge is the inherent fragility of the center-proximity correspondence method, as its accuracy is entirely dependent on the quality of the segmentation output. The method for establishing lesion correspondence across time points, a center-proximity assignment rule, is conceptually straightforward but operationally fragile. This approach is a cascade of three components: lesion centroid propagation via registration, segmentation of the resulting VOI, and assignment of the lesion ID to the segmented component nearest to the VOI's center. Its effectiveness is therefore critically dependent on the tandem success of accurate registration and correct segmentation. As demonstrated in the previous section, the segmentation step is sensitive to the misalignments introduced by registration errors.

The impact of this dependency is clear when comparing the correspondence outcomes between the baseline (Figure \ref{fig:lesion_assignment_a}) and follow-up (Figure \ref{fig:lesion_assignment_b}) scans. We observe a significant drop in correct assignments at follow-up. While a substantial fraction of this shift is due to an increase in true negatives, cases where a lesion has resolved and the model correctly segments nothing, the proportion of outright failures also increases. The combined rate of false assignments and false negatives, which represent clinically critical tracking errors, rises from an already high 26.3\% at baseline to 31.8\% at follow-up. In a clinical setting, where response assessment depends on the consistent tracking of specific lesions, an error rate affecting nearly one-third of follow-up measurements is untenable. Such failures introduce a significant risk of incorrect patient staging and flawed evaluations of treatment efficacy.

The controlled displacement experiment directly illustrates the mechanism behind this failure. As shown in Figure \ref{fig:correspondence_impact_shift}, as a lesion is artificially shifted further from the VOI center, the rate of incorrect assignments and particularly false negatives increases progressively. This trend is directly related to the segmentation behavior observed in Figure \ref{fig:dice_vs_distance}. Once a lesion is displaced beyond a certain threshold, the model fails to segment it. This results in no connected component being available for the correspondence algorithm, leading inevitably to a false negative. This cascading effect, where registration error leads to segmentation failure, which in turn causes correspondence failure, exposes a fundamental vulnerability in this approach.

\section{Conclusion}
\label{sec:conclusion}
In this study, we evaluated the ULS23 single time-point segmentation model in a longitudinal context, revealing critical vulnerabilities when applying such isolated analysis tools to serial scans. In follow-up cases, we found the model's performance to be fragile and highly dependent on registration accuracy to satisfy its assumption of a centered lesion. As our experiments demonstrated, moderate misalignments, common in propagated coordinates, frequently lead to complete segmentation failure. This directly undermines our correspondence method, resulting in a high rate of incorrect or missed lesion associations in follow-up scans.

While the model's center bias could be addressed by training with a more varied spatial distribution of lesions, we argue the fundamental issue is the use of cascaded, independent steps which leads to compounding errors. The more robust path forward is to develop unified, end-to-end frameworks that are inherently longitudinal. Recent work has already started to move in this direction, extending nnUNet to integrate the temporal dimension by concatenating multi-timepoint images as additional input channels.  \cite{TDW_rokuss_2024,nnUNet_isensee_2021,LesionLocator_rokuss_2025} . Future efforts should elaborate on this, and focus on models that jointly optimize for anatomically and temporally consistent lesion correspondence, moving away from chaining single-timepoint tools towards building truly time-aware systems for oncological tracking.

\end{document}